# Quantum-Like Behavior of Nonlinear Classical Oscillator

S.A. Rashkovskiy

*Institute for Problems in Mechanics of the Russian Academy of Sciences,
101/1 Vernadskogo Ave. Moscow, 119526, Russia
E-mail: rash@hotbox.ru*

**Abstract.** We construct the classical dynamical system which has a quantum-like behavior. We have shown that the energy-time uncertainty relation takes place for the system and it has purely classical nature. We investigate the behavior of the system and propose a "classical" explanation of Franck-Hertz experiments.

**Keywords:** classical dynamical system; quantum-like behavior; Franck-Hertz experiment, energy-time uncertainty relation

**PACS:** 03.65.Sq, 03.65.Ta, 05.45.-a, 45.20.D-, 45.50.-j

## INTRODUCTION

In the course of development of quantum mechanics the point of view was formed [1] that it is impossible to construct the classical mechanical system which would possess by discrete energy levels and "correct quantum behavior", i.e. which would be able to undergo a transition spontaneously from one stationary level to another, radiating discrete amount of energy - quanta.

For example, [1] gives the explanation why quantum system differs from classical one: «…energy of a system of classical particles is an essentially continuous quantity: *no matter how one modifies the force law, not even by introducing additional dynamical variables, can one change this situation: the fact that the energy of a system of particles is limited to a discrete set of allowed values is a result which falls outside Classical Mechanics…*».

Despite popular opinion that quantum systems have no classical analogues, the classical systems possessing quantum-like behavior have been recently found and investigated [2-9]. These systems are purely classical ones, but they have a number of quantum properties. A feature of these systems is their essential nonlinearity. These facts show that not only classical systems can have quantum properties, but also the quantum systems can be described by the classical models, reproducing a number of important properties of quantum systems.

In this paper we generalize and investigate the model [9] and we show that well-known energy-time uncertainty relation is a natural result of the model.

# ATOM AS A DYNAMICAL SYSTEM

Let us briefly list well-known properties of atoms:
(i) Atom has a discrete set of stationary states. The stationary state with a least energy is the ground state; all other stationary states are excited ones. (ii) It is impossible to disturb the atom from the ground state even under the enough strong action (the Franck-Hertz experiment). (iii) If the atom is in an excited state it can undergo a transition spontaneously to a lower energy level. This transition is accompanied by emission of the energy quanta. (iv) Spontaneous transition of atom to a stationary state with higher energy is impossible. Such transition can occur only forcedly. This stimulated transition is accompanied by absorption of corresponding energy quanta by the atom. All of these properties of atom are described in detail by quantum mechanics and quantum electrodynamics.

The goal of this work is to build one-dimensional classical dynamical system which has above mentioned properties. Despite we consider one-dimensional system we will call it conditionally "an atom".

Let us postulate the expected properties of such a system.

(a) Atom is a non-linear classical dynamical system: the state of the system is described by coordinate, $q$, and momentum, $p$. Energy of the system at any instant is described by Hamilton function $H(p,q)$. (b) The system has a discrete set of stationary states with energies $E_n$. Only states that satisfy the Bohr-Sommerfeld quantization rule [1]

$$J = \frac{1}{2\pi} \iint_{H(p,q) \leq E} dp\,dq = \left(n + \frac{1}{2}\right)\hbar \tag{1}$$

are stationary ones, where $n = 0, 1, 2, \ldots$, $E$ is the system energy. (c) Being at a stationary state the atom is a Hamiltonian system:

$$\dot{q} = \frac{\partial H(p,q)}{\partial p}, \quad \dot{p} = -\frac{\partial H(p,q)}{\partial q} \tag{2}$$

(d) The ground state $E_0$ is absolutely stable; it can be considered as an attractor of the system.
(e) The excited states ($n > 0$) are absolutely unstable; they can be considered as the saddle points of dynamical system. (f) The energy of the system plays a role of Lyapunov function.

It is easy to build such a system. For example the dissipative system

$$\dot{q} = \frac{\partial H(p,q)}{\partial p}, \quad \dot{p} = -\frac{\partial H(p,q)}{\partial q} - \mu(J)\frac{\partial H(p,q)}{\partial p} \tag{3}$$

satisfies the above mentioned properties if function $\mu(J)$ satisfies the conditions: (i) $\mu(J) \geq 0$ for $J \geq \frac{1}{2}\hbar$; (ii) $\mu(J) < 0$ for $J < \frac{1}{2}\hbar$; (iii) the discrete set of values (1) are the solutions of equation $\mu(J) = 0$. This is evident from analysis of derivatives of energy with respect to time

$$\frac{dH}{dt} = -\mu(J)\left(\frac{\partial H(p,q)}{\partial p}\right)^2 \leq 0 \qquad (4)$$

Equation (4) shows that decrease in energy is the result of dissipation of energy in the system under consideration, but from point of view of transitions of atom, this loss of energy can be considered as a radiation of energy, for example in the form of electromagnetic pulse (a quantum). For example, the function $\mu(J) = a\left(J - \frac{\hbar}{2}\right)\cos^2(\pi J/\hbar)$ satisfies the above mentioned requirements; where $a > 0$ is any function or constant.

Consider next the harmonic oscillator as an example of such a system. In this case

$$J = E/\omega, \quad H(p,q) = \frac{1}{2m}\left(p^2 + m^2\omega^2 q^2\right) \qquad (5)$$

and equation of motion (3) is as follows [9]

$$\ddot{q} + \omega^2 q = -\frac{\gamma}{\hbar}\left(E - \frac{\hbar\omega}{2}\right)\dot{q}\cos^2(\pi E/\hbar\omega) \qquad (6)$$

where $m$ is the mass of particle, $\omega$ is the circular frequency of oscillator, $\gamma > 0$ is the non-dimensional parameter; we will show that it should be considered as variable one in order to fit the results of this model with well-known results of quantum mechanics.

## ENERGY-TIME UNCERTAINTY RELATION

The equation of changing of energy (4) takes the form

$$\frac{dE}{dt} = -\frac{\gamma m}{\hbar}\left(E - \frac{\hbar\omega}{2}\right)\dot{q}^2\cos^2(\pi E/\hbar\omega) \qquad (7)$$

For small deviation $\xi = E_n - E \ll E_n$ from a stationary state, equation (6) is reduced to

$$\frac{d\xi}{dt} = \frac{\gamma m}{\hbar}\left(E_n - \frac{\hbar\omega}{2}\right)\dot{q}^2(\pi\xi/\hbar\omega)^2 \qquad (8)$$

Solution of the equation (8) can be represented in the form

$$\Delta E \Delta t = \hbar \frac{1}{\gamma\pi^2\left(E_n/\hbar\omega - \frac{1}{2}\right)(E_n/\hbar\omega)} \qquad (9)$$

where $\Delta E$ is an initial disturbance in energy; $\Delta t$ is a characteristic transition time.

Expression (9) takes the form of well-known energy-time uncertainty relation

$$\Delta E \Delta t \geq \hbar \qquad (10)$$

if we assume $\gamma\pi^2(E/\hbar\omega)^2 = \gamma_0 \sim 1$.

The solution (10) has a very simple and evident sense: the more initial disturbance in energy the less duration of transition. Contrary to quantum mechanics, the energy-time uncertainty relation (10) for system under consideration has purely classical nature.

In this case equation of motion (6) takes the form

$$\ddot{q}+\omega^2 q = -\frac{\gamma_0}{\hbar\pi^2}\left(\frac{\hbar\omega}{E}\right)^2\left(E-\frac{\hbar\omega}{2}\right)\dot{q}\cos^2(\pi E/\hbar\omega)+f(t) \quad (11)$$

where $f(t)$ describes the external disturbances which can stimulate the transitions.

## RESULTS OF CALCULATIONS

The equation (11) has been solved numerically. We use the non-dimensional variables

$$\omega t \to t,\ (m\omega/\hbar)^{1/2}q \to q,\ E/\hbar\omega \to E = \frac{1}{2}(\dot{q}^2+q^2) \quad (12)$$

and equation (11) takes the form

$$\ddot{q}+q = -\frac{2\gamma_0}{\pi^2}\frac{\dot{q}}{(\dot{q}^2+q^2)^2}(\dot{q}^2+q^2-1)\cos^2\left(\frac{\pi}{2}(\dot{q}^2+q^2)\right)+f(t) \quad (13)$$

The equation (13) was solved for different initial conditions $\dot{q}(0)=V_0$ at $q(0)=0$ and $\frac{2\gamma_0}{\pi^2}=0.2$. We consider periodic disturbances $f(t)=A_0\sin\Omega t$, where $A_0$, $\Omega$ are the constants.

Calculations show that the system eventually undergoes a transition into one of the stationary states (1) at any initial conditions.

The "occupancy" of the energy levels at different times as a function of initial velocity $V_0$ is shown at Fig.1. One can see that for $t<10$, a multitude of initial values of velocity $V_0$ correspond to the same stationary state. This means that the particle can reach the same stationary energy level having the different initial velocity.

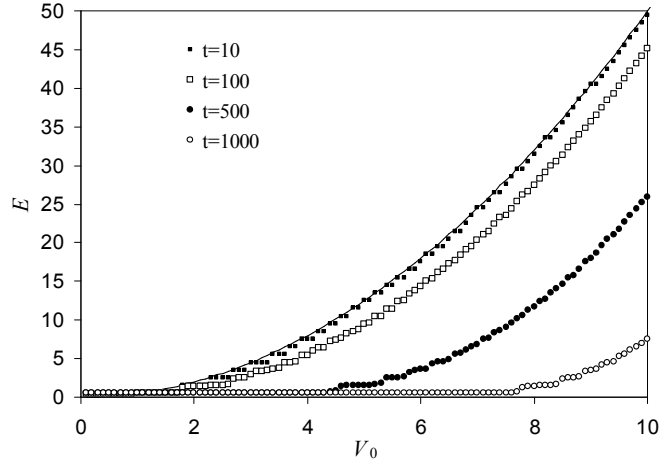

**FIGURE 1.** "Occupancy" of quantum levels for different initial velocities: the markers are solutions of the equation for different times, the line is the classical dependence $E=\frac{1}{2}V_0^2$; $A_0$=0.05; $\Omega$=2

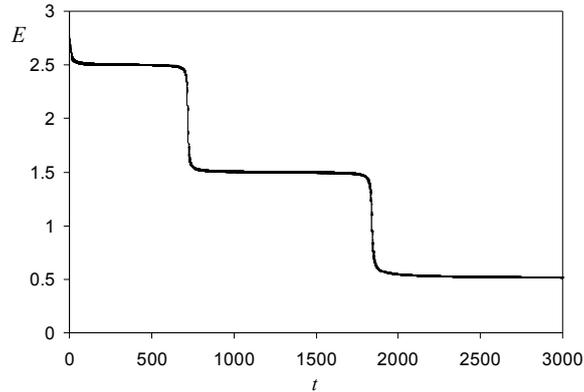

**FIGURE 2.** Dependence of energy of the system on time; $A_0$ =0.05; $\Omega$ =2

The energy of oscillator is approximately defined by the classical expression $E = \frac{1}{2}V_0^2$ at high values of quantum numbers $n \gg 1$. This result can be considered as a manifestation of Bohr's correspondence principle.

Due to disturbances, the excited states of the system have a finite lifetime and system undergoes a transition spontaneously to a lower energy level. One can see that a peculiar "fall of the levels" occurs: the system existing in the excited states eventually undergoes a transition to a lower energy level. As a result of this process the excited states with higher energy are devastated while the lower energy levels are filled up.

Calculations show that the system always undergoes a transition to the ground state, independently of initial conditions. Lifetime of the levels is different for different levels: the higher energy the less lifetime.

The examples of calculations for relatively long time of the process are shown in Fig. 2. It shows that the system stays at one of the excited states for some time but eventually it undergoes a transition spontaneously to a level with a lower energy. Duration of the transition is of order of few periods of oscillations of oscillator. It means, that energy, emitted by the system, is not "spread" on time, but represents a compact portion - a quantum.

## FRANCK-HERTZ EXPERIMENT

Franck-Hertz experiment is one of the fundamental experiments in quantum physics. It is widely believed that the Franck-Hertz experiment cannot be explained within the limits of the classical mechanics. Let us show that a "classical" explanation of Franck-Hertz experiments can be given within the limits of the classical model under consideration.

Let us consider a qualitative explanation of the Franck-Hertz experiments [9]. We will use two notions: collision and scattering. Collision means a direct mechanical collision of free electron with oscillator while a scattering means a final result of the process of electron-oscillator interaction as a whole. Let us imagine that a free moving electron collides with the oscillator (in a real experiment with atom). For simplicity we will assume, that this collision is absolutely elastic. As a result of the collision the momentum and energy of particles are changed abruptly in accordance with usual laws of classical mechanics. Moreover, let us assume that this collision is instant: before and after the collision the oscillator and electron do not interact directly.

We start from the case when the electron energy is less than the difference of the energies between the first excited state and the ground state of oscillator. As a result of this collision the oscillator leaves the ground state but does not reach the first excited state. In accordance with dynamical equation (13) the oscillator tends to come back to the ground state under the action of "quantum friction". Energy losses during this process can be considered as electromagnetic radiation which the oscillator emits when it comes back to the ground state. In the absence of free electron when disturbances in the oscillator were caused by other reasons, it can be assumed that energy emitted by oscillator is simply dissipated in environment. In case of interaction of oscillator with free electron we assume that the energy emitted by the oscillator in its motion to the ground state is absorbed by free electron. If it is assumed that all energy emitted by the oscillator is absorbed by free electron, then, eventually, free electron will return itself all the energy lost at collision with the oscillator and the scattering of electron will be elastic.

Let us consider now a case when the energy of the free electron is higher than the difference of energies of the first excited and ground states of oscillator.

In this case as a result of collision the oscillator undergoes a transition to a state with higher energy than energy of the first excited state. In accordance with oscillator dynamics described by equation (13), as it was shown above, the oscillator undergoes a transition spontaneously to the first excited state and can remains here for a relatively long time without emitting the energy received in the collision with the electron. During the lifetime of first excited state of oscillator the free electron have enough time to fly away and it "does not gain back its energy". Thus inelastic scattering of electron takes place in this case.

Detailed description of the algorithm of calculations of Franck-Hertz experiment within the limits of the model under consideration can be found in [9].

The dependences of mean velocity and mean energy of free electrons after scattering on initial electron energy obtained in the calculations are shown in Fig.3. The mean velocity of electron in these calculations is similar to the electric current measured in Franck-Hertz experiment. The character of the curves in Fig. 3 is in qualitative agreement with the experimental data obtained from the Franck-Hertz experiment with different gases.

## CONCLUSION

Contrary to popular belief, we have shown that it is possible to construct a classical dynamical system which possesses a specific set of properties which are characteristic for actual quantum objects. The proposed model does not pretend to give an alternative description of quantum systems because it does not take into account the wave-particle duality which is a fundamental property of quantum objects.
Such models may be useful for semi-classical simulations of actual systems with taking into account the spontaneous and stimulated transitions, interactions of atoms with radiation (absorption and scattering of photons) etc. The method developed can be easily extended to hydrogen atom and more complex quantum systems.

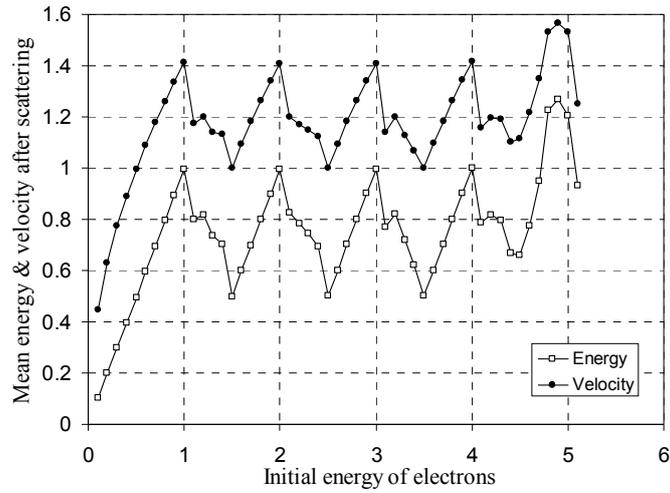

**FIGURE 3.** Dependencies of mean energy and mean velocity of free electrons after scattering on initial energy of electrons